# Characterizing the temperature dependence of Fano-Feshbach resonances of Ultracold Polarized Thulium

V. A Khlebnikov[1], V. V Tsyganok[1,2,3], D.A. Pershin[1,2], E. T Davletov[1,2], E. Kuznetsova[4], A.V. Akimov[1,5,6]

[1]Russian Quantum Center, Business center "Ural", 100A Novaya str., Skolkovo, Moscow, 143025, Russia

[2]Moscow Institute of Physics and Technology, Institutskii per. 9, Dolgoprudny, Moscow region, 141700, Russia.

[3]National University of Science and Technology MISIS, Leninsky Prospekt 4, Moscow, 119049, Russia,

[4]Nizhny Novgorod State University, 23 Gagarin Street, Nizhny Novgorod, 603950, Russia.

[5]PN Lebedev Institute RAS, Leninsky prospekt 53, Moscow, 119991, Russia.

[6]Texas A&M University, TAMU 4242, College Station, TX 77843, USA

email: akimov@physics.tamu.edu

# I. ABSTRACT

Recent studies demonstrated anomalous temperature shifts for some Fano-Feshbach resonances of thulium atoms. These anomalies were explained by the variation in light intensity in the optical dipole trap, which accompanied changes in temperature. In addition, a temperature-related transformation of the statistics of the interresonance spacing was demonstrated [1]. Here, we analyze the shifts of isolated s- and d-type Fano-Feshbach resonances of ultracold thulium atoms with temperature for a fixed depth of an optical dipole trap. The measurements are consistent with the 3-body recombination-based theory of the temperature-related resonance shift and enable the extraction of the resonance parameters, particularly the magnetic moments of closed channel states. This parameter and the known polarizability of the open channel enable us to separate the contributions of the temperature and Stark shift to the overall shift of the resonances and show the dominant role of the Stark effect in the overall shift.

# II. INTRODUCTION

Highly magnetic atoms, such as Cr, Er, Dy and Ho [2–5], have recently attracted great attention because they can interact via anisotropic and long-range magnetic dipole-dipole interactions. The dipole-dipole interaction causes many novel phenomena, such as the modification of the Bose Einstein Condensate (BEC), including deformation and d-wave collapse [6–8], observation of dipolar ferrofluid [9], highly dense Feshbach resonances with chaotic distribution statistics [10], quantum liquid crystals and other exotic quantum many-body phases [11,12], supersolid phase in quantum gases [13–15].

Thulium with an unfilled 4f shell and a magnetic moment of 4µB belongs to the family of highly magnetic atoms and is currently actively studied. In particular, our group has demonstrated a notably dense spectrum of Fano-Feshbach resonances with temperature-dependent statistics [1] and condensed thulium atoms to the BEC state [16]. In this work, we continue exploring an ultracold thulium gas and study the low-field Feshbach resonances of s- and d-types in the collisions of ultracold thulium atoms.

Fano-Feshbach (FF) resonances have become a conventional method to tune the interaction strength in low-energy atomic collisions [17]. The scattering length resonantly increases when the energy of colliding particles in an entrance channel matches the energy of a bound molecular state in a closed channel. Resonance is enabled by tuning the energies of the entrance and closed channels by an external magnetic or electric field via Zeeman or Stark effects, respectively. The FF resonances manifest themselves in an enhanced loss of atoms from the trap at specific magnetic fields. The magnetic field-dependent losses are usually explained by the 3-body recombination, whose rate increases near an FF resonance [18–20]. In particular, as was discussed for the case of erbium [21], the increased loss rate of atoms from the optical dipole trap near the center of the FF is caused by the 3-body recombination, which is also the case for the fermionic mixture described in [20].

Previously, our group obtained detailed loss spectra for thulium in low magnetic fields (up to 24 G) in a temperature range of $2-12\,\mu K$ [1]. Similar to other lanthanides [21], two types of resonances are observed: the so-called s-type resonances, which already exist at the lowest temperatures of ~ 2 µK and are associated with s-collisions in the entrance channel, and the d-type resonances, which emerge at higher temperatures and are associated with d-type collisions in the entrance channel [21]. In the experiment, thulium is polarized to the lowest energy Zeeman state $|F=4,\ m_F=-4\rangle$ and captured in an optical dipole trap [1]. The theory also predicts that the maxima of both s- and d-type resonances shift to higher magnetic fields with increasing temperature due to an increase in mean energy of colliding atoms. An increase in mean atomic energy requires larger magnetic fields to equalize the energy of the closed and open channels. Surprisingly, some of the resonances for thulium were found to shift to smaller magnetic fields with increasing temperature, which contradicts the previously developed theory.

The negative shift can be explained by the change in temperature and intensity of the optical dipole trap in our experiment, which can result in different ac-Stark shifts of the entrance and closed channels and consequently an additional (negative) shift of the resonance position. An optically induced shift of a magnetic Feshbach resonance was observed in two-body collisions [22] but not in three-body collisional spectra. In our case, the sizeable ac-Stark shift of both entrance and closed channels can be due to a notably large tensor component of polarizability [23] at the wavelength of our optical dipole trap (ODT) (532 nm), which is close to the 530.7 nm transition line of thulium. The present work aims to study in more detail the FF resonance shift dependence on both temperature and trap light intensity.

## III. EXPERIMENT

Thulium is an element from the lanthanide series, which includes actively studied Er [3,24,25], Dy [4,26–28] and Ho [5,29]. Thulium has a complicated electronic structure due to a partially filled $4f$ electronic shell submerged below a closed 6s shell, which induces the electronic ground state with large orbital and total angular momenta ($F = 4$ for thulium). The large orbital angular momentum leads to significant anisotropy in short-range van der Waals and dipolar interactions due to the large magnetic moment of the ground state of $4\mu_B$ [30]. These anisotropies (mostly van der Waals anisotropies) are responsible for the notably dense and even chaotic spectrum of FF resonances [21] in comparison to alkali metals or chromium.

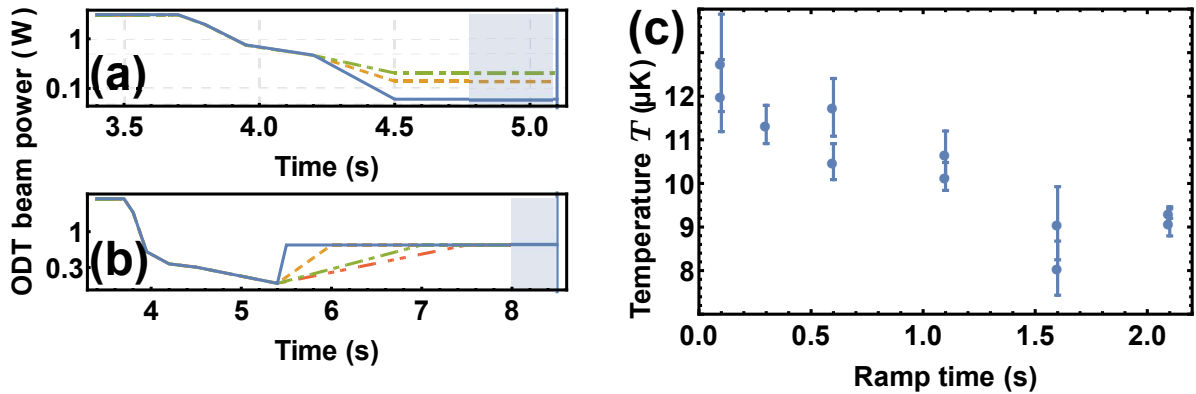

*Figure 1 A – ODT ramps in the experiment to reach the desired temperatures: power-dependent ODT depth. B – same for fixed ODT depth. The shaded areas indicate when the scanned magnetic field was on. C – temperature after various durations of a linear ramp of light power in the horizontal ODT beam with fixed depth ramp, indicated at the panel B. The delay between the beginning of the ramp and the temperature measurement time is identical for all ramps.*

The details of the experimental setup are shown at [1]. Typically, the temperature of the atomic cloud is strongly correlated with the depth of the optical dipole trap (ODT) [31]. To obtain a better understanding of the nature of the resonance shift and perform a theoretical analysis of the observed shifts in the framework of the 3-body recombination theory, the position of a resonance was measured at a constant depth of the ODT, which is at constant light power of the trapping beam. While experimentally measured parameter is ODT power $P$, in calculations it is more convenient to use intensity $I$ and the depth of ODT $U$ can be found as:

$$I = \frac{2}{\pi w_x w_y} \cdot P$$
$$U = I \cdot \frac{\operatorname{Re}(\alpha_{tot})}{2\varepsilon_0 c} = \frac{2}{\pi w_x w_y} \cdot \frac{\operatorname{Re}(\alpha_{tot})}{2\varepsilon_0 c} \cdot P \tag{1}$$

with beam waists values: $w_x = 16\ \mu\mathrm{m}$ and $w_y = 26\ \mu\mathrm{m}$ measured as described at [23]. Values of scalar and tensor polarizabilities of thulium in the light of our ODT are provided at [23]. Due to presence of large tensor and vector polarizability calculation of $\alpha_{tot}$ requires knowledge of the geometry of the experiment. In our case the ODT beam had horizontally oriented linear light polarization and was orthogonal to the vertically oriented scanning and storing magnetic field leading to $\alpha_{tot} = 620$ a.u..

As a first step, the cooling sequence in Figure 1A,B was realized: after a precooling step, the optical power was monotonously reduced to perform the regular evaporation sequence (Figure 1A), or after reaching a sufficiently low temperature ($5\ \mu\mathrm{K}$), the power was increased to a higher value (Figure 1B). The selection of the ODT rise time defined the final temperature of the atomic ensemble (Figure 1C). Using this technique, we could vary the temperature of the atomic cloud in the range of $8-14\ \mu\mathrm{K}$ while maintaining a constant depth of the ODT. Each FF resonance was measured by detecting the number of atoms that remained in the trap after a $500 ms$ exposure time (see Figure 2B) using the absorption imaging technique. In this study, two resonances were selected due to their relative isolation from other FF resonances [1]: one of the s-type near the magnetic field of 4.3 G and another of the d-type near the magnetic field of 3.6 G (Figure 2C).

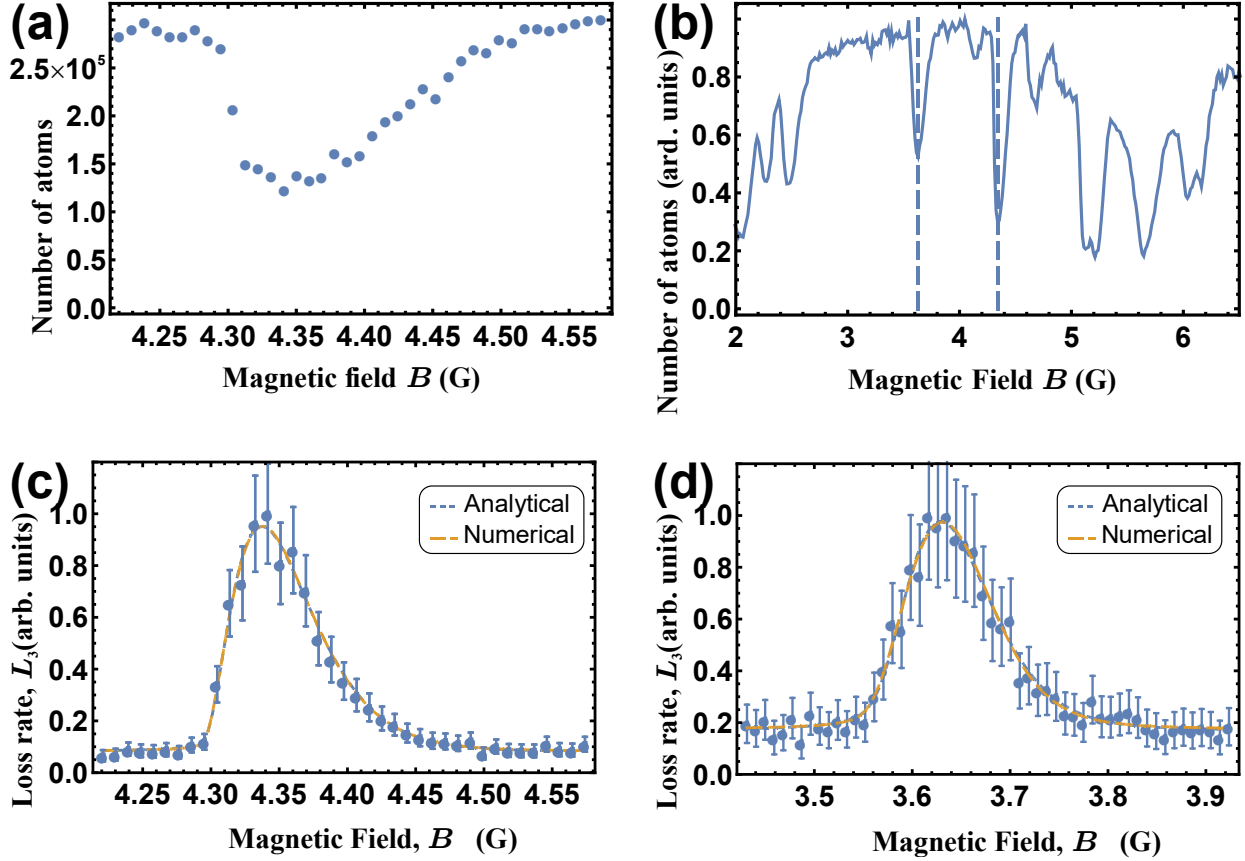

Figure 2 A – An example of the FF resonance profile. B – FF spectral fragment, which indicates the selected FF resonances for this study. C – Loss rate coefficient $L_3(B,T)$ for the s-resonance at 4.3 G. The fit was performed by analytical formula (6) (blue dotted line) and the numerical integration of the full model (3) (orange dashed line). D – Loss rate coefficient $L_3(B,T)$ for the d-resonance at 3.63 G. The fit was performed by the analytical formula (6) (blue dotted line) and numerical integration of the full model (3) (orange dashed line). Plots C and D are taken at 13 μK.

## IV. THEORETICAL ANALYSIS

### A. The model

In this study, thulium atoms are polarized to the lowest-energy Zeeman state $|F=4, m_F=-4\rangle$, which eliminates possible dipolar relaxation [32]. We analyze the atomic loss profile $N(T,B)$ to extract the 3-body loss rate coefficient $L_3(B,T)$, which is related to the number of atoms in the case of pure 3-body losses as follows [21]:

$$L_3(B,T) = \frac{1}{6n_0^2(T)\tau}\left(\left(\frac{N_0}{N(T,B)}\right)^2 - 1\right), \tag{2}$$

where $n_0(T) = N_0 \bar{\nu}^3 \left(\frac{\pi m}{k_B T}\right)^{3/2}$ is the average initial atomic density, $N_0$ is the number of atoms at the beginning of the measurements, $T$ is the temperature of the atomic ensemble, and $\tau$ is the exposure time. Temperature was measured at the moment when the scanning field was switched on at least twice – before and after the magnetic field scan; therefore, the temperature measurements contain both statistical and fit errors. The number of atoms $N_0$ at the beginning of the magnetic field scan was also extracted from these measurements. The mean trap frequency $\bar{\nu}$ was measured by the method described in [1]. The loss rate $L_3(B,T)$ profiles obtained from the atom loss profiles (e.g., Figure 2A) for s- and d-resonances with their errors are presented in Figure 2C,D.

Compared to our previous work [1], the profile of the loss rate coefficient was directly fitted by the expression for $L_3(B,T)$, which is obtained in the coupled-channel description of the 3-body recombination [21]:

$$L_3(T,B) \propto \frac{a}{T^3}\int_0^C dE_3 E_3^{\lambda+2} e^{-E_3/T} \frac{1}{\left[E_3 - \delta\mu(B-B_0)\right]^2 + \left[\Gamma_{br} + A_\lambda E_3^{\lambda+2}\right]^2} + b \tag{3}$$

where the relative energy of the 3 atoms $E_3$ is measured in temperature units; $\delta\mu$ is the difference of magnetic moments of the closed and entrance channels; $\Gamma_{br}$ is the rate of trimer breakup into a Feshbach molecule and an atom; $A_\lambda$ is related to the energy width of the entrance channel $\Gamma(E_3) = A_\lambda E_3^{\lambda+2}$, where $\lambda = 0$ and $\lambda = 2$ for the s- and d-type entrance channels with the least repulsive centrifugal barriers; $B_0$ is the magnetic field corresponding to the recombination resonance. Parameters $\delta\mu$, $B_0$, $\Gamma_{br}$, $A_\lambda$, a and b of the model were varied during a fit for each resonance at many temperatures. Afterwards, they are averaged. $a$ and $b$ are associated with the experimental conditions. If $b$ is a positive constant due to single-particle losses, other possible losses can also affect amplitude $a$. The difference $\delta\mu$ of magnetic moments of the closed and entrance channels is positive, since the incoming atoms are in their ground magnetic state. The position of the resonance maximum depends on the temperature. A positive shift $\Delta B$ of the resonance maximum is expected, since the mean energy of atoms in an entrance channel increases with temperature, so a larger magnetic field is required to match the energies of closed and entrance channels and fulfill the resonance condition.

The fitting by expression (3) does not enable us to determine parameter $A_\lambda$ in the available temperature range. While fitting data sets for a particular resonance at different temperatures, the values of $A_\lambda$ vary by several orders of magnitude, but the term $A_\lambda E_3^{\lambda+2}$ in the denominator of the Lorentzian in (3) is always much smaller than $\Gamma_{br}$ (see the APPENDIX for the numerical calculation details).

A simplified approximation can be obtained if one neglects the second term in the denominator of the quasi-Lorentzian part of the integrand in (3), i.e., when $A_\lambda E_3^{\lambda+2}$ and $\Gamma_{br}$ are much smaller than $T$ in the range of $E_3$, where the integrand remains nonnegligible. In this limit, the Lorentzian function can be treated as a delta function $\delta\left(E_3 - \delta\mu(B-B_0)\right)$, so the integral is easily calculated, and the expression for $L_3(B,T)$ becomes:

$$L_3(T,B) = \frac{\tilde{a}}{T^3} \delta\mu^{\lambda+2} (B-B_0)^{\lambda+2} e^{-\frac{\delta\mu(B-B_0)}{T}} + b \tag{4}$$

From (4), it follows that the position of maximum 3-body losses changes with temperature as follows (the same limit was described in [21]):

$$B = B_0 + (\lambda+2)\frac{T}{\delta\mu} \tag{5}$$

and the maximum value of $L_3(B,T)$ scales with temperature as $T^{\lambda-1}$. Although this model for $L_3(B,T)$ fits the data well, if we use it to extract $B_0$ and $\delta\mu$ (5), their values differ from those directly obtained from model (4). This discrepancy cannot be settled without bringing back at least one parameter that is responsible for the width of the resonance.

Integral (3) with $A_\lambda = 0$ has an analytical expression. One can represent the Lorentzian as a sum of two terms and take expression 2.3.6.13 from [33]. In this case, the analytical expression for $L_3(B,T)$ is:

$$L_3(T,B) = \frac{\widetilde{Amp}}{T^3} i \left[ e^{\frac{-\delta\mu(B-B_0)+\frac{i\Gamma_{br}}{2}}{T}} \left(-\delta\mu(B-B_0)+\frac{i\Gamma_{br}}{2}\right)^{\lambda+2} \times \right.$$

$$\times \Gamma\left(-\lambda-2, \frac{-\delta\mu(B-B_0)+\frac{i\Gamma_{br}}{2}}{T}\right) - e^{\frac{-\delta\mu(B-B_0)-\frac{i\Gamma_{br}}{2}}{T}} \left(-\delta\mu(B-B_0)-\frac{i\Gamma_{br}}{2}\right)^{\lambda+2} \times \quad (6)$$

$$\left. \times \Gamma\left(-\lambda-2, \frac{-\delta\mu(B-B_0)-\frac{i\Gamma_{br}}{2}}{T}\right)\right] + const$$

where $i$ is the imaginary unit, $\Gamma(\alpha,z)$ is the upper incomplete gamma-function, and Amp is a constant. The dependence of the maximum loss on the temperature is now more complicated. When the temperature is much smaller than $\Gamma_{br}$, the position of maximal losses shifts with temperature as:

$$B = B_0 + (\lambda+3)\frac{T}{\delta\mu} \quad (7)$$

(the maximum value of $L_3(B,T)$ now scales with temperature as $T^\lambda$). However, at high temperatures when $\Gamma_{br}$ is negligible, linear dependence (5) returns, except now the $B_0(T=0)$ position does not coincide with $B_0$ from the $L_3(B,T)$ model.

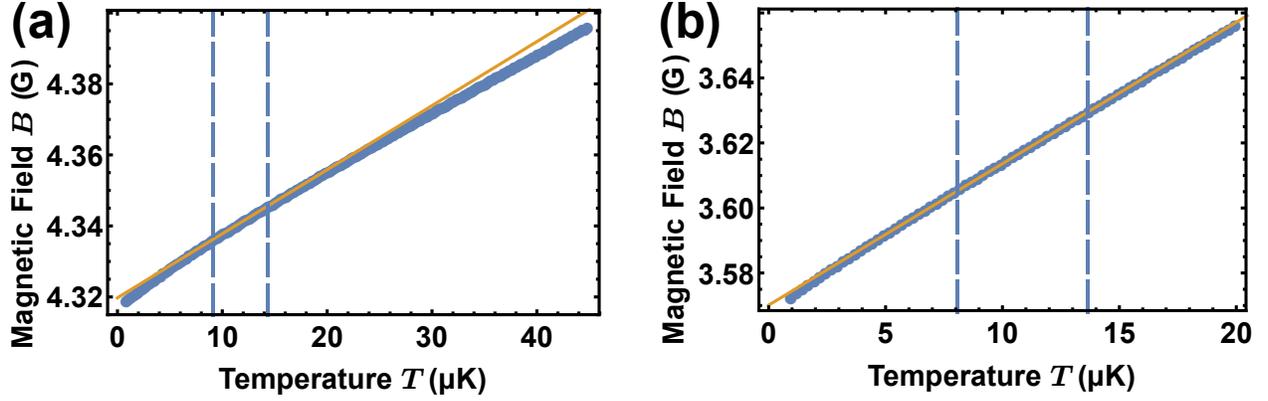

*Figure 3 Calculation of the slope of the temperature dependence of the resonance position for averaged values of the fit parameters obtained from fit profiles (thick blue line) and its approximation with the straight line (orange curve). Only the experimentally available temperature range (vertical dashed lines) is used to obtain coefficient $\xi_\lambda$ for (8). A – An example of such fitting for the s-resonance at 4.3 G; B – An example of such fitting for the d-resonance at 3.6 G.*

### B.  The Fitting procedure

After fitting the profiles with model (6), we extract and average parameters $B_0, \delta\mu, \Gamma_{br}$ and use them to numerically model the resonance positions in the experimental range of temperatures to determine slope $\xi_\lambda$ assuming a linear dependence of the form (see Figure 3):

$$B = \tilde{B}_0 + \xi_\lambda \frac{T}{\delta\mu} \qquad (8)$$

In the experimental range of temperatures (between black vertical lines), the linear dependence is a good approximation. We use expression (8) assuming that $\delta\mu$ and $\tilde{B}_0$ are again unknown to fit the temperature shift of the resonance position obtained from (6). The new value of $\tilde{\delta\mu}$ obtained from the fit is now incorporated into (6) as a fixed number, and the procedure is repeated until $\tilde{\delta\mu}$, which is used to fit the profiles, coincides with that obtained from (8). This procedure converges in a few steps for both s- and d-type resonances.

We investigate the temperature behavior of two resonances: s-resonance at $4.3G$ and d-resonance at $3.6G$. After fitting the $L_3(B,T)$ profile by (6) with $\lambda = 0$ for the s-resonance and $\lambda = 2$ for the d-resonance at several temperatures, we obtain the following averaged values for the model parameters: $\delta\mu = 570 \pm 30 \mu K/G$, $B_0 = 4.293 \pm 0.002 G$, $\Gamma_{br} = 2.7 \pm 2.5 \mu K$, and the value of prefactor $\xi_0$ from (8) is 2.001 for the s-resonance; $\delta\mu = 690 \pm 100 \mu K/G$, $B_0 = 3.55 \pm 0.01 G$, $\Gamma_{br} = 15 \pm 11 \mu K$ and the value of prefactor $\xi_2$ is 4.018±0.001 for the d-resonance. In both cases, we are actually very close to the negligible $\Gamma_{br}$ approximation ($\xi_\lambda \approx (\lambda + 2)$ as should be in this case. See (5)). However, if we fit the positions of maximal by expression (8) using the above values of prefactor $\xi_\lambda$, $\delta\mu$ and $B_0$ differ significantly from those obtained by directly fitting by (6). Namely, for the s-resonance, we obtain $\delta\mu = 1220 \pm 410$ $\mu K/G$ and $B_0 = 4.32 \pm 0.01 G$; for the d-resonance, $\delta\mu = 870 \pm 150$ $\mu K/G$ and $B_0 = 3.57 \pm 0.01 G$. If we now use the latter values of $\delta\mu$ as fixed numbers in model (6), we obtain the following parameters: $\delta\mu = 1220 \mu K/G$, $B_0 = 4.316 \pm 0.002 G$, $\Gamma_{br} = 56 \pm 8 \mu K$, and the value of the prefactor from (8) is $\xi_0 = 2.254 \pm 0.004$ for the s-resonance; $\delta\mu = 870 \mu K/G$, $B_0 = 3.563 \pm 0.002 G$, $\Gamma_{br} = 33 \pm 6 \mu K$ and the value of the prefactor is $\xi_2 = 4.075 \pm 0.002$ for the d-resonance. Fitting the new positions of maximal losses by (8) gives: for the s-resonance, $\delta\mu = 1480 \pm 700$ $\mu K/G$ and $B_0 = 4.32 \pm 0.01 G$; for the d-resonance, $\delta\mu = 890 \pm 160 \mu K/G$ and $B_0 = 3.57 \pm 0.01 G$. Thus, the values of $\delta\mu$ obtained from model (6) coincide within errors with the corresponding values obtained from the linear dependence (8).

The parameters stop changing after only 2 iterations (Tables T1 and T2) and acquire their final values: $\delta\mu = 1430 \pm 630 \mu K/G$, $B_0 = 4.320 \pm 0.002 G$ and $\Gamma_{br} = 73 \pm 10 \mu K$ for the s-resonance and $\delta\mu = 890 \pm 160 \mu K/G$, $B_0 = 3.564 \pm 0.002 G$ and $\Gamma_{br} = 35 \pm 6 \mu K$ for the d-resonance. The positions of maximal losses were numerically obtained from model (6). The $L_3(B, T)$ magnitudes at the positions of maximal losses and the positions are presented in Figure 4

*Table 1 Evolution of the parameter values for the 4.3G s-resonance at the iterative procedure described in the text*

|  | δμ [μK] from loss profile | $B_0$ [G] from loss profile | $\Gamma_{br}$ [μK] from loss profile | $\xi_0$ | $B_0$ [G] from (8) | δμ [μK] from fitting shift of data | $B_0$ [G] from fitting shift of data |
|---|---|---|---|---|---|---|---|
| 0th Iter | 570±30 | 4.293±0.002 | 2.7±2.5 | 2.001 | 4.294 | 1220±410 | 4.32±0.01 |
| 1st Iter | 1220 | 4.316±0.002 | 56±8 | 2.25 | 4.319 | 1480±700 | 4.32±0.01 |
| 2nd Iter | 1480 | 4.320±0.002 | 77±10 | 2.36±0.01 | 4.323 | 1430±620 | 4.32±0.01 |
| 3rd Iter | 1430 | 4.320±0.002 | 73±10 | 2.34±0.01 | 4.322 | 1430±630 | 4.32±0.01 |

*Table 2 Evolution of parameter values for the 3.6 G d-resonance at the iterative procedure described in the text*

|  | δμ [μK] from loss profile | $B_0$ [G] from loss profile | $\Gamma_{br}$ [μK] from loss profile | $\xi_0$ | $B_0$ [G] from (8) | δμ [μK] from fitting shift of data | $B_0$ [G] from fitting shift of data |
|---|---|---|---|---|---|---|---|
| 0th Iter | 690±100 | 3.55±0.01 | 15±11 | 4.018±0.001 | 3.552 | 870±150 | 3.57±0.01 |
| 1st Iter | 870 | 3.563±0.002 | 33±6 | 4.075±0.002 | 3.566 | 890±160 | 3.57±0.01 |
| 2nd Iter | 890 | 3.564±0.002 | 35±6 | 4.084±0.002 | 3.568 | 890±160 | 3.57±0.01 |

### C. Discussion

When both temperature and trap laser intensity change, we observe a negative shift $\Delta B < 0$ with increasing temperature [1]. The negative shift can be explained by the difference in Stark shifts of the entrance and closed scattering channels, which results in the dependence of the resonant magnetic field $B_0$ on the trap laser intensity.

The Stark effect can be the origin of the negative shift of the resonance position because of a large tensor component of thulium polarizability in the trap operating at the 532 nm wavelength [23]. While the resonance still shifts with the temperature, as it should at the absence of the different polarizabilities of the open and closed channels, it will also shift due to the fact, that open and close channel shift differently with the electric field. Indeed, since open and closed channel have to have different magnetic quantum numbers and therefore, due to tensor polarizability, have to shift differently with Electric field. Thus, the resonance, which matches kinetic energy in the open channel with energy of the bound state shifts with electric field of ODT. If so, the position of the resonance acquires an additional shift and becomes:

$$\Delta B = \frac{1}{2\varepsilon_0 c} \frac{\delta \alpha}{\delta \mu} I + \xi_\lambda \frac{k_B T}{\delta \mu} \tag{9}$$

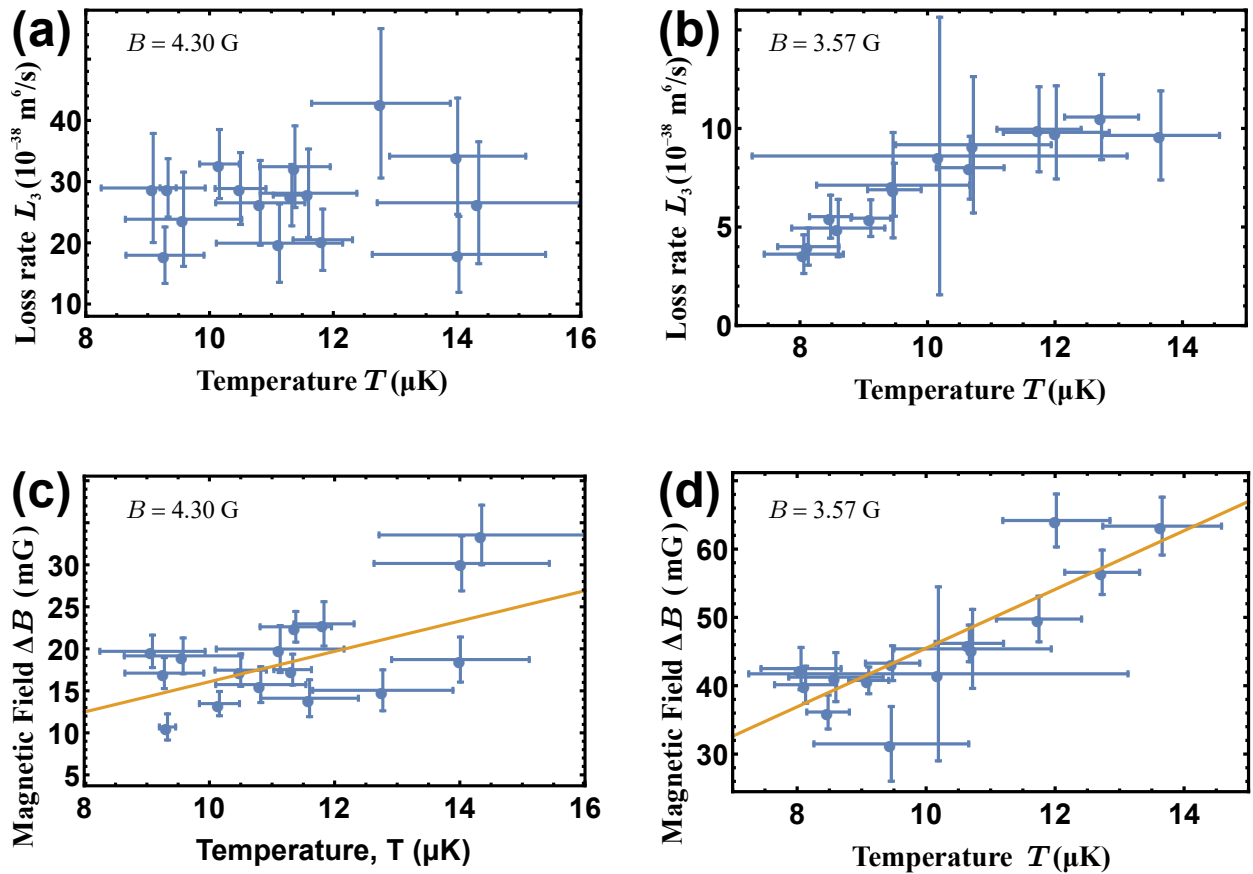

Figure 4. Temperature dependence of the maximal 3-body loss rate for the (A) s-type Fano-Feshbach resonance at 4.30 G and (B) d-type resonance at 3.6 G. (C)

*Temperature shift of the position of the maximum loss for s-type Fano-Feshbach resonances. The magnetic field of 4.303 G is taken as the origin on the plot. The solid line represents the fit with parameters $\delta\mu = 1430 \pm 630\,\mu K/G$, $B_0 = 4.320 \pm 0.002 G$. (D) Shifts of positions of d-type resonances versus temperature. The magnetic field of 3.569 G is taken as the origin on the plot. The solid line represents the fit with parameters $\delta\mu = 890 \pm 160\,\mu K/G$, $B_0 = 3.564 \pm 0.002 G$.*

Here, $\varepsilon_0$ is the vacuum permittivity, $c$ is the speed of light, $I$ is the laser intensity, and $\delta\alpha = \alpha_{cl} - \alpha_{op}$ is the difference in polarizability for closed and entrance channels with $\alpha_{op} = 3\alpha$, where $\alpha$ is the polarizability of thulium at a 532-nm trap wavelength [23]. The shift becomes negative if $\alpha_{cl} < \alpha_{op} - 2\varepsilon_0 c \xi_\lambda k_B \dfrac{\Delta T}{\Delta I}$.

Having obtained the magnetic moment of the closed channel during measurements with a constant trap laser intensity, we can estimate the polarizability of this channel using expression (9) and resonance positions (Figure 5C,D), which were obtained at various ODT depths by fitting the $L_3(B,T)$ profile with model (6) and the fixed $\delta\mu$ value. The $T(I)$ dependence for the data sets from Figure 5C,D is known from the experiment (Figure 5B) and linear. Hence, after the depth of the trap increases, there was sufficient time to bring the system to equilibrium, and the losses are mostly due to Feshbach resonance enhanced collisions instead of evaporation. Thus, taking $\delta\mu = 1430\,\mu K/G$ (or equivalently $\delta\mu = 21.3\,\mu_B$) for the s-resonance to fit the $L_3(B,T)$ loss profiles, we obtain the resonance shift and extract $\delta\alpha = -280 \pm 90\,a.u.$ using $\xi_0 = 2.34$. Similarly, for the d-resonance with $\delta\mu = 890\,\mu K/G$ ($\delta\mu = 13.2\,\mu_B$) and $\xi_2 = 4.084$, we obtain $\delta\alpha = -235 \pm 52$ a.u. To convert the polarizability from atomic to S.I. units in which the equation (9) is written one should multiply it by $4\pi\varepsilon_0 a_B^3$ where $a_B$ is Bohr radius.

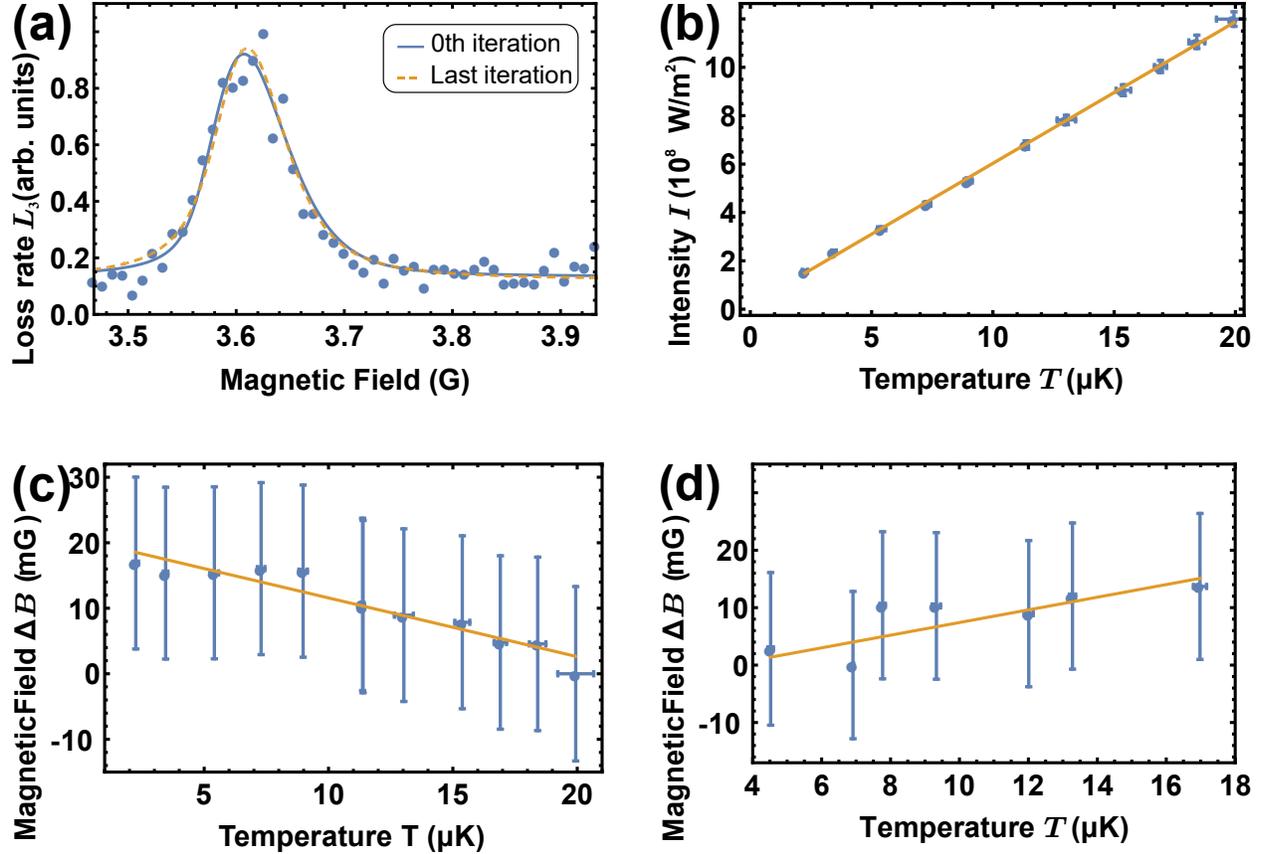

*Figure 5 A – Example of data fitting with expression (6), where the blue line is the fit with all parameters being kept free (i.e., the 0th iteration). The orange dashed line is the fit at the last iteration, where $\delta\mu$ is fixed by the value from expression (8) implemented to the obtained position shifts in the previous iteration. The temperature at which the resonance was recorded is $8.6\ \mu K$. B – Intensity vs. temperature plot for experiments with traps of different depths. C – Shift of the maximum loss position for the 4.3 G s resonance at various intensities of trap light. The effective difference (if the Stark effect is not considered) in the magnetic moments of the closed and entrance channels is given in the legend. D – the same as C but for the d-resonance at 3.6 G.*

Finally, we observed a negative shift in Fano-Feshbach resonances with increasing temperature and trap light intensity. We associate it with a strong Stark effect in our optical dipole trap and confirm this statement by data obtained at a constant trap depth (intensity of trap laser). The positions of resonances were obtained in the framework of the 3-body recombination theory, which enabled us to extract the zero-temperature position of a resonance and its closed channel magnetic moment. The strong dependence of the resonance position of resonances on the trap light intensity enables us to determine the polarizability of a closed channel.

Thus, thulium is a favorable element for the studies of Bose-Einstein condensate-related phenomena. The easy tunability of thulium-thulium interactions in low fields using Fano-Feshbach resonances makes it an interesting candidate for quantum simulations. It is especially exciting that the resonance parameters can be tuned with both magnetic and optical fields by adding an additional knob to manipulate the system. Moreover, optical control enables the manipulation of the system on short length and time scales, which adds more flexibility to the experiment.

## V. ACKNOWLEDGMENTS

This study was supported by Russian Science Foundation grant #18-12-00266. We acknowledge support from Rosatom.

## APPENDIX

Integral (3) was numerically calculated. Therefore, it was necessary to set the upper limit of the integral, which was selected as $\delta\mu\Delta B + 5\Gamma_{br} + 2(\lambda+2)T$ with $\Delta B = 0.1\,\text{G}$ for the d-resonance and $\Delta B = 0.25\,\text{G}$ for the s-resonance (see Figure 2 with the loss profiles). The validity of the selected cutoff limit for the integral was checked for several magnetic field values within $\Delta B$ from the position of maximum losses. For all tested magnetic fields, the numerical integration result does not change by more than $10^{-4}\%$ if the cutoff value for the upper limit for the integral (3) is taken twice as large as the described value.

The estimation of error bars was performed as follows. For the first three columns in Table 1 and Table 2, the errors were estimated as the errors of the fitting procedure for the loss profiles with model (6). For the last two columns, the error accounts for the error of the linear model fit for the positions of maxima of losses versus temperature and systematic errors in measurements of the temperature and magnetic field. Relation (8) is also used to relate the errors in the determination of the maximum and temperature uncertainties.